\documentclass[useAMS,usenatbib, usegraphicx]{mn2e}
\usepackage{epsfig}
\usepackage{epstopdf}
\usepackage{graphicx}
\usepackage{amsmath}
\usepackage{amssymb}
\usepackage{natbib}
\include{epsf}
\newcommand{\kms}{\ensuremath{{\rm km\,s}^{-1}}}
\newcommand{\mpc}{\ensuremath{{\rm Mpc}}}
\newcommand{\msun}{\ensuremath{M_{\odot}}}
\newcommand{\edc}{\ensuremath{\epsilon_{DC}}}
\newcommand{\fstar}{\ensuremath{f_{\star}}}
\newcommand{\infinity}{{\infty}}
\newcommand{\apj}{ApJ}
\newcommand{\apjl}{ApJ}
\newcommand{\mnras}{MNRAS}

\newcommand{\apjs}{ApJS}
\newcommand{\nat}{Nat}

\newcommand{\aap}{A\&A}

\newcommand{\physrep}{Physics Reports}

\begin{document}

\title{Galaxy Statistics in Pencil-beam Surveys at High Redshifts}

\author[Mu{\~n}oz, Trac, and Loeb]{
Joseph A. Mu{\~n}oz\thanks{E-mail:jamunoz@cfa.harvard.edu}, 
Hy Trac, and 
Abraham Loeb\\Harvard-Smithsonian Center for Astrophysics, 60 Garden St., MS 10, Cambridge, MA 02138, USA
}
\maketitle

\begin{abstract}
Surveys of faint galaxies at high redshifts often result in a
``pencil-beam'' geometry that is much longer along the line-of-sight
than across the sky.  We explore the effects of this geometry on the
abundance and clustering of Lyman-break galaxies (LBGs) and
Lyman-$\alpha$ emitters (LAEs) in current and future surveys based on cosmological N-body simulations
which adequately describe the nonlinear growth of structure on small
scales and compare to linear theory.  We find that the probability distribution of the LBG abundance
is skewed toward low values since the narrow transverse dimension of
the survey is more likely to probe underdense regions. Over a range
that spans 1--2 orders of magnitude in galaxy luminosities, the
variance in the number of objects differs from the commonly used
analytic prediction and is not dominated by Poisson noise.
Additionally, nonlinear bias on small scales results in a
one-dimensional power spectrum of LAEs using a James Webb Space
Telescope field-of-view that is relatively flat, markedly different from the expectation of linear perturbation theory.  We discuss how these results may affect attempts to measure the UV background at high redshifts, estimate the relationship between halo
mass and galaxy luminosity, and probe reionization by measuring the
power-modulating effect of ionized regions.
\end{abstract}


\section{Introduction}

Over the past few years, searches for Lyman-break galaxies (LBGs) and
Lyman-alpha emitters (LAE) have revealed new populations of young,
star-forming objects at redshift beyond $z=6$ \citep[e.g.][]{Stark07,
Richard08, Bouwens08a, Bouwens08b} that shed light on the star
formation history at high redshifts and the sources responsible for
cosmic reionization.  In addition, evidence has been found for a
population of massive, evolved systems around $z=5$ \citep{Mobasher05,
Wiklind08} that hints at an earlier period of higher star formation
than has yet to be seen in surveys of bright LBGs and LAEs.

Theoretically, it is expected that the bulk of the star formation
during reionization had taken place in less luminous galaxies than
previously observed\citep[e.g.][]{BL01, WL06}.  Dwarf galaxies are supposed to
supply the bulk of the radiation that reionizes the Universe as well
as provide the building blocks of the Milky Way and the possible birth
place of globular clusters \citep[e.g.][]{KG05, GK06, Moore06, Madau08b,
Munoz09}.  Attempts to probe the faint end of the luminosity function
have often sacrificed field-of-view for the sake of higher flux
sensitivity.  Objects in the $11.2\, \rm{arcmin}^2$ Hubble Ultra Deep
Field (HUDF), for example, can be seen in the $z_{850,{\rm AB}}$ passband down
to an apparent magnitude of about $29$.  In an even more extreme case,
a single long-slit spectroscopic survey for gravitationally-lensed
LAEs has a field-of-view of only about $32\, \rm{arcsec}^2$ (which is
even smaller in the unlensed source plane) but can achieve large
boosts in sensitivity if positioned on the gravitational lensing
critical line of a foreground galaxy cluster.  The James Webb Space
Telescope (JWST\footnote{http://www.jwst.nasa.gov/}), planned for
launch in 2013, will have a better sensitivity, enabling observers to
probe ``pencil beams'' out to even higher redshifts.

Recent data analysis from deep and narrow galaxy surveys has begun to be
incorporated into theoretical models of structure formation and
reionization.  Several studies have
attempted to relate the observed UV and Lyman-alpha luminosities from
these sources to the masses of the dark matter halos that host them by
fitting results from simulations and semi-analytic
prescriptions to the observations and so understand the clustering and abundance of such galaxies \citep[e.g.][]{LeDelliou05, LeDelliou06, SLE07, KTN07, Nagamine08, Orsi08, KTN09}.  In addition, it has been shown that cosmic reionization in the high-redshift Universe can be probed by observing its effects on the galaxy power spectrum \citep[e.g.][]{dBL06, McQuinn07, WL07}.

In this paper, we study the underlying abundance and clustering
statistics of halos in narrow fields-of-view. Our goal is to relate
the theory of structure formation to the large amount of observational
data from high-redshift galaxy surveys.  We take into account the
effect of large-scale density fluctuations as well as the
nonlinearities on small scales.  We use state-of-the-art numerical
simulations to examine the field-to-field sample distribution and
one-dimensional power spectrum of ``pencil-beam" galaxy surveys that
are much longer along the line-of-sight than across the field-of-view.
Our analysis is designed to help observers in the interpretation of
related observational data in the future.  

While, similar studies of the cosmic variance and count distributions in high-redshift dropout surveys were carried out \citep[e.g.][]{TS08, Overzier09}, our analysis differs from \citet{TS08} in that we use our simulated halo catalogs to create mock surveys viewed along the light-cone and probes more deeply than both studies since our simulations can resolve halos as small as $\sim 10^{8}\,\msun$.

Our paper is organized as follows.  In \S \ref{sec:sim} we describe
our simulations.  The probability distributions of object counts in
high-redshift pencil surveys are analyzed in \S \ref{sec:pdist}.  In
\S \ref{sec:1Dpower}, we explore the one-dimensional power spectrum
obtained from such surveys, the nonlinear bias, and the usefulness of
such measures in relating galaxy luminosity and halo mass.  Finally,
we discuss our results and their implications for observations in \S
\ref{sec:conclusions}.

\section{The Simulation Data}\label{sec:sim}

To simulate observations of high-redshift galaxies, we adopt halo
catalogs from an N-body simulation and semi-analytically dress each
halo with a luminosity based on its mass.  These catalogs were
produced by the simulation used in \citet{Trac08} which resolved halos
with masses as low as $\sim 10^{8}\,\msun$, almost two orders-of-magnitude smaller than those resolved in simulations by \citet{TS08}, in a box of comoving size
$143\,\mpc$ (assuming a Hubble constant of $H_0=70. {\rm km/s/Mpc}$).  This approach allows us to probe the faintest
high-redshift galaxies within a sufficiently large simulation box that
includes the important large-scale fluctuations.  A flat, $\Lambda$CDM
cosmology was assumed with cosmological parameters from the {\it
{Wilkinson Microwave Anisotropy Probe}} 5-year data release \citep{Komatsu09, Dunkley09}, $\left(
\Omega_{m}, \Omega_{\Lambda}, \Omega_{b}, h, \sigma_8,
n_s\right) = \left( 0.28, 0.72, 0.046, 0.70, 0.82,0.96\right)$.

We determined a UV or Lyman-$\alpha$ luminosity for each halo in the
simulation according to the prescription of \citet{SLE07} using their simple assumption that each halo hosts no more than a single bright galaxy.  Their model for star formation assumes a fraction, $\fstar$, of the baryons in a halo with mean baryon fraction forms stars in an amount of time given by the cosmic age at the observed redshift times the duty cycle $\edc$.  The efficiency is assumed to be constant with mass for halos with circular velocity above $100\,\kms$, however, star formation in smaller halos is suppressed so that the efficiency is a power-law with mass to the $2/3$ power.  The resulting star formation rate is converted to a UV luminosity (for LBGs) according to the \citet{MPD98} model or to a Lyman-$\alpha$ luminosity (for LAEs) by assuming a fixed fraction, $T_{\alpha}$, of the $4\times10^{53}$ ionizing photons emitted per star formation rate in ${\rm \msun/yr}$ (for a Salpeter IMF) escape the galaxy and the IGM.  For a particular type of galaxy, luminosity $L$ is associated with halo mass $M_{\rm halo}$ so that the abundance of observed systems with luminosity $L$ is equal to the \citet{ST99} abundance of halos with mass $M$ times the duty cycle.  We adopt the \citet{SLE07} best-fit values of $(\edc,\fstar)=(0.25,0.16)$ and $(1.0,0.063)$ for LBGs and LAEs, respectively.  Since $\edc < 1$ for LBGs, we
appropriately consider only a fraction, $\edc$, of them to be observed
in each snapshot of a galaxy survey.  

Finally, since we hope to present results that are not too model-dependent, we will continue to refer to both the galaxy luminosity and halo mass in the calculations that follow and in relevant figures.  In this way, future models deemed more appropriate may be applied to our results for a given halo mass.

\section{Probability Distribution of Galaxy Counts}\label{sec:pdist}

\begin{figure}
\begin{center}
\includegraphics[width=\columnwidth]{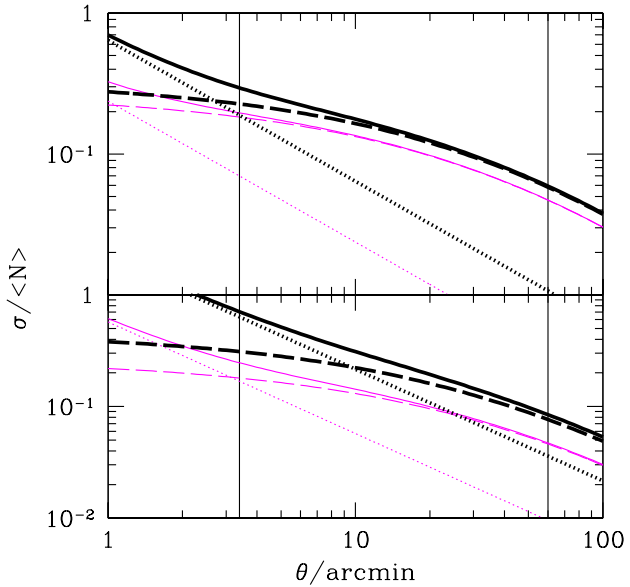}
\caption{\label{fig:sigfov} The contributions to the total variance (Eq. \ref{eq:var}; solid lines) in LBG dropout surveys as a sum of cosmic variance (dashed lines) and Poisson shot noise (dotted lines) contributions (i.e. the first and second terms, respectively, on the right-hand-side of Eq. \ref{eq:var}).  The top and bottom panels show results for surveys extending from z=6-8 and z=8-10, respectively.  Thin lines assume a luminosity threshold of $z_{850,{\rm AB}}$=29, while for thick ones, the cut is at $z_{850,{\rm AB}}$=27.
}
\end{center}
\end{figure}

\begin{figure}
\begin{center}
\includegraphics[width=\columnwidth]{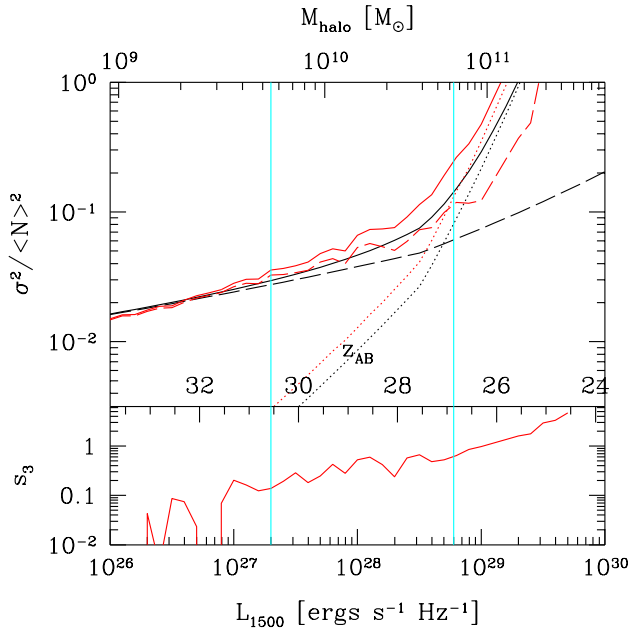}
\caption{\label{fig:var6-8} The upper panel shows relative
contributions to the fractional variance in the number counts of
galaxies as a function of UV luminosity, z-band magnitude, or host
halo mass in counts of LBGs in a dropout survey spanning the redshift
interval $z=6$--$8$ with a $3.4'\times3.4'$ field-of-view.  Solid lines
show the total variance, while long-dashed and dotted lines show the
contributions from sample variance and Poisson noise, respectively.
Red curves show simulation results, while black curves were calculated
analytically.  Vertical lines brackett the region where the variance
is higher than expected due to the skewness of the full count
probability distribution but is not Poisson-dominated.  The lower
panel shows the skewness of the full galaxy count probability
distribution calculated from the simulation (Eq.~\ref{eq:skew}).  }
\end{center}
\end{figure}

In this section, we create mock samples of LBGs in pencil beams surveys
at different redshifts.  Since photometric redshifts of LBGs are
typically imprecise, we consider groups of these galaxies in the
redshift intervals of $z=6$--$8$ and $z=8$--$10$.  Many LBGs already have
been observed in the HUDF around $z=6$ \citep[e.g.][]{BI06, Bouwens06}, but
observers have not yet had the same success at higher redshifts
\citep[e.g.][]{Bouwens08a, Bouwens08b}.  Given its higher sensitivity, it is
hoped that JWST will be able to observe LBGs at these redshifts.

To construct our samples of LBGs, we stitch together simulation
slices in the given redshift range so that each survey volume is
viewed along the light-cone.  This allows the mass function of halos
to evolve along the line-of-sight, as needed to
represent high-redshift surveys and to correctly analyze their
statistics \citep{ML08b}.  For simplicity, we have assumed a top-hat
selection function over each redshift interval.

The time spacing between simulation snapshots is $10\,{\rm Myr}$.  This means that there are $29$ and $16$ simulation slices between $z=6-8$ and $z=8-10$, respectively.  The comoving sizes of these slices vary from $21.1\,\mpc$ at $z=6$ to $27.3\,\mpc$ at $z=8$ and $32.6\,\mpc$ at $z=10$.  We include in our $z=6-8$ lightcone catalog halos from a slice $21\,\mpc$ thick at one end of the simulation snapshot at $z=6$, those from the adjacent slice $21\,\mpc$ thick in the snapshot taken $10\,{\rm Myr}$ earlier, and so on until we include the halos in a slice $27\,\mpc$ thick at $z=8$.  The $z=8-10$ lightcone catalog is produced similarly.

The comoving distances within the redshift intervasl $z=6-8$ and $z=8-10$ are about $700$ and $500\,\mpc$, respectively, while the simulation box is only $143\,\mpc$ on a side.  In the process of stitching together slices, when we come to the end of the simulation box, we simply rotate the box and again stitching together slices through the box.  We make the rotation by giving new coordinates $(x'_1,x'_2,x'_3)$ to each halo with previous position $(x_1,x_2,x_3)$ such that $(x'_1,x'_2,x'_3)=(x_2,x_3,x_1)$; we also make the additional rotation $(x'_1,x'_2,x'_3)=(x_1,x_3,x_2)$ every third time we run through the simulation box so that there is no repetition in skewers along the line-of-sight.  These rotations suppress power in the survey on the scale of the simulation length.

For each redshift range, we calculate the field-to-field variance in the
count of LBGs in a ``pencil beam" of a given field-of-view using both linear theory and from the simulation data.  The total variance in a survey is the sum of contributions from cosmic variance and Poisson shot noise \citep[see, e.g.,][]{Somerville04}:
\begin{equation}\label{eq:var}
\frac{\sigma^2_{\rm tot}}{\left<N\right>^2}=\frac{\left<N^2\right>-\left<N\right>^2}{\left<N\right>^2}=\frac{\sigma^2_{\rm hh}}{\left<N\right>^2}+\frac{1}{\left<N\right>},
\end{equation}
where $\left<N\right>$ is the mean number of LBGs in each skewer.  The cosmic (or sample) variance (the first term on the right hand side of Eq.~\ref{eq:var}) results from the field sometimes lying in a region of high density with a lot of structure due to the clustering of galaxies and sometime lying in an under-dense region or a void.  This contribution can be calculated from linear perturbation theory (i.e. based on the linear power-spectrum, $P(k)$) as a function of the minimum halo mass, $M$, hosting observed galaxies as
\begin{equation}\label{eq:cosvar}
\sigma^2_{\rm hh}(M)=\frac{(b_{eff}(M)\,D(z))^2}{(2\,\pi)^3}\int\,\,P(k)\,W^2_{xyz}\,d^3\vec{k},
\end{equation}
where $W_{xyz}=W(k_x)\,W(k_y)\,W(k_w)$, $k=\sqrt{k_x^2+k_y^2+k_w^2}$,
$D(z)$ is the linear growth factor evaluated, for simplicity, at the
midpoint of the given redshift range, and $b_{eff}$ is the
\citet{SMT01} linear bias integrated over all masses above $M$ and
weighted by the \cite[][hearafter ST]{ST99} halo mass function
\citep{Matarrese97,ML08b}.  The window function, $W(k_i)$, is the
Fourier transform of a top-hat in the $i$-th dimension and is given
by:
\begin{equation}\label{eq:window}
W(k_i)=\frac{\sin(k_i\,a_i/2)}{k_i\,a_i/2}.
\end{equation}
In equation~(\ref{eq:window}), $a_x$ and $a_y$ are the narrow dimensions
of the skewer-shaped survey evaluated at $z$.  The subscript $w$
refers to coordinates along the line-of-sight.  The length of the
survey, $a_w$, is given by the comoving distance between the redshift range of interest.

To calculate the survey statistics, we take $\theta=3.4'$ (the field-of-view the HUDF and approximately of JWST) and count the number of LBGs in each of the $256$ ($225$) possible skewer positions perpendicular to the front face of our $z=6-8$ ($z=8-10$) lightcone simulation volume.  While nearby skewers may be correlated, distant skewers are slightly anticorrelated.  Taking each possible skewer exactly once is equivalent to choosing an infinite number of randomly located skewers since each skewer position will be sampled the same number of times on average.  Moreover, given that we expect correlations on the size of the simulation box to be negligible, there is no difference between this statistical measurement and one with only a single skewer and multiple realizations of the simulation volume are produced using different initial conditions.

\begin{figure}
\begin{center}
\includegraphics[width=\columnwidth]{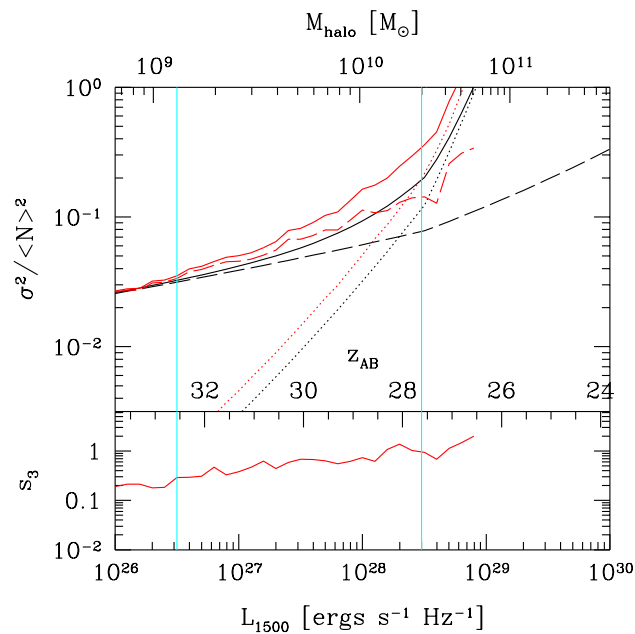}
\caption{\label{fig:var8-10} Same as Figure \ref{fig:var6-8} but for
the redshift interval $z=8$--$10$.}
\end{center}
\end{figure}

Figure \ref{fig:sigfov} compares the contributions from cosmic and Poisson variance as calculated by linear theory (Eq. \ref{eq:var} and \ref{eq:cosvar}) for $a_x=a_y$ as a function of the opening angle of the survey, $\theta=a_x/\chi(z)$.  This diagram can be used to calculate the effectiveness of future surveys with large fields of view.  Given the limiting size of our simulation volume, a sufficient number of skewers is not obtainable to perform comparable numerical calculations for large surveys.  However, figures \ref{fig:var6-8} and \ref{fig:var8-10} show the agreement for the HUDF field-of-view.

\begin{figure}
\begin{center}
\includegraphics[width=\columnwidth]{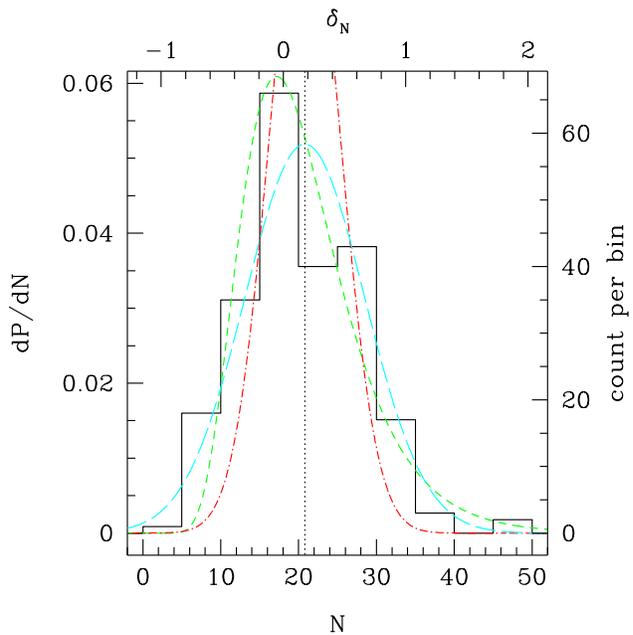}
\caption{\label{fig:pdist8-10} The full probability distribution for
the number $N$ of LBGs in a dropout survey spanning the redshift
interval $z=8$--$10$ with a minimum galaxy luminosity of $10^{28}
\rm{ergs/s/Hz}$.  The solid black curve shows the distribution as a
histogram measured from the simulation, while the red dot-dashed, cyan
long-dashed, and green short-dashed lines overlay Poisson, Gaussian,
and log-normal distributions with the same mean, mean and standard
deviation, and log-mean and log-standard deviation, respectively.  The
vertical, dotted line denotes the mean number of objects in the survey
over many pointings.  The distributions are plotted as differential
probability in a differential abundance bin on the left, vertical
axis as a function of the measured abundance of LBGs, $N$, on the
lower, horizontal axis and as a function of the overdensity in the
abundance $\delta_N=N/\left<N\right>-1$ on the upper, horizontal axis.
Additionally, the counts of pointings per abundance bin for the
simulation histogram is shown on the right, vertical axis.  These
counts per bin show that Poisson errors in the simulation histogram
are smaller than the differences between the overlaid distributions.
The simulation results match most closely with the skewed log-normal
distribution.  }
\end{center}
\end{figure}

Figures \ref{fig:var6-8} and \ref{fig:var8-10} compare the contributions to the variance in the simulations with calculations based on linear theory for the $z=6-8$ and $z=8-10$ ranges, respectively.  They
suggest that the observed statistics are well approximated by the
analytic calculations at the low luminosity limits.  However, while linear theory predicts a Gaussian probability distribution of the count of halos with variance given by equation~\ref{eq:cosvar}, the simulated probability distribution of bright LBGs has a non-Gaussian shape.  This can be clearly seen by eye in Figure
\ref{fig:pdist8-10}, where the solid histogram representing the
measured probability distribution is best represented by a log-normal
distribution.  The skewness as a function of minimum luminosity is
presented in the bottom panels of figures \ref{fig:var6-8} and
\ref{fig:var8-10}.  We define the skewness as the third moment of the
probability distribution normalized by the variance to the $3/2$
power:
\begin{equation}\label{eq:skew}
s_3=\frac{\left<\left(N-\left<N\right>\right)^3\right>}{\left(\sigma^{2}\right)^{3/2}}.
\end{equation}
The seemingly large amplitude variations in the skewness at low
luminosity for $z=6$--$8$ are due to small numerical fluctuations
around the nearly zero skewness from the simulation, plotted on a log
scale.  Deviations between the analytic and simulation values of the
sample variance grow as the skewness becomes more significant.  This
behavior is a manifestation of nonlinear clustering on the small
scales probed by the narrowness of the skewer.

Finally, the figures show sharp regimes where the field-to-field
variance is dominated either by cosmic or Poisson variance.  For the
two redshift ranges $z=6$--$8$ and $z=8$--$10$ this division comes at around
$L\left(M=5\times10^{10}\msun\right)=6\times10^{28}\,\rm{ergs/s/Hz}$ and
$L\left(M=2\times10^{10}\msun\right)=3\times10^{28}\,\rm{ergs/s/Hz}$,
respectively.  The nonlinearity at higher luminosities causes the
sample variance to dominate at masses which are $\sim 50$--$100\%$
larger than otherwise expected.  However, since the Poisson variance
is unaffected by nonlinearities, there remains only a specific
mass range (shown in the figures) over which the total variance is
different than expected.

\section{Line-of-sight (One-Dimensional) Power Spectrum}\label{sec:1Dpower}

Next, we consider the one-dimensional (1D) galaxy power spectrum
measured along the line-of-sight of a ``pencil-beam" survey at high
redshift with the same $3.4'\times3.4'$ field-of-view considered in
the previous section.  We explore what the information along the
radial direction can or cannot tell us about the nature of
high-redshift galaxies given the narrow field-of-view. We also examine
whether or not it can be used to distinguish between luminous galaxies
hosted by different mass halos, and ultimately to probe reionization.
We are considering the ``galaxy" power spectrum in the sense that we
apply a duty cycle for LAEs and translate halo masses into hosted
Lyman-$\alpha$ luminosities, but our results do not include the
effects of reionization on the power spectrum that we expect to
observe in real surveys \citep[e.g.][]{dBL06, McQuinn07, WL07}.  

\begin{figure}
\begin{center}
\includegraphics[width=\columnwidth]{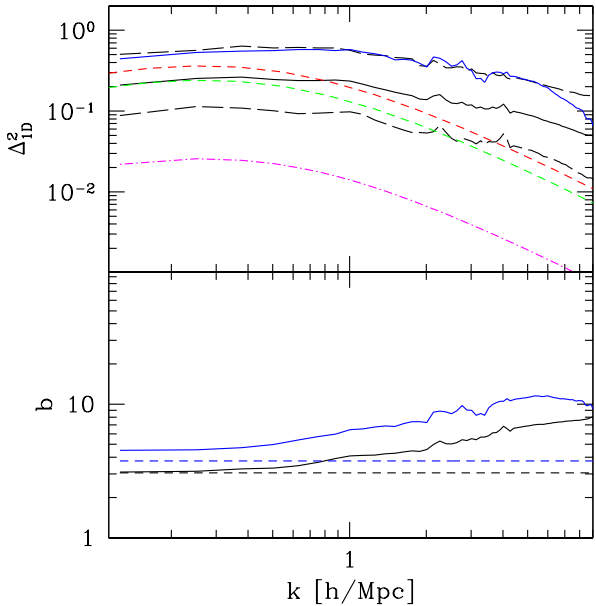}
\caption{\label{fig:bias1D_z6} The 1D power spectrum of halos hosting
LAEs in a survey with a $3.4'\times3.4'$ field-of-view at $z=6$.  The
black (blue) lines represent values for a minimum Lyman-$\alpha$
luminosity of $L_{\rm L\alpha,min}=10^{40}\,(10^{41})\,\rm{ergs/s}$
corresponding to $M_{\rm halo}=1.6\times10^9\,(6.4\times10^9)\,\msun$.  In
the upper panel, solid lines show the average log of the
Poisson-subtracted, dimensionless power spectrum.  Long-dashed lines
indicate the estimated $1-\sigma$ variation in the log of the
amplitude of the power spectrum from skewer-to-skewer.  For clarity,
these lines only appear for the lower luminosity threshold, but they
are about the same size for the higher luminosity one.  Short-dashed,
green and red lines show the linear theory predictions for the
dimensionless halo power spectrum given the lower and higher
luminosity thresholds, respectively, while the dotted-dashed, magenta
curve shows the linear theory calculations for the dark matter.  The
lower panel shows the nonlinear bias (Eq.~\ref{eq:bias}) measured from
the simulation (solid lines) and the linear, ST bias (short-dashed
lines).  }
\end{center}
\end{figure}

\begin{figure}
\begin{center}
\includegraphics[width=\columnwidth]{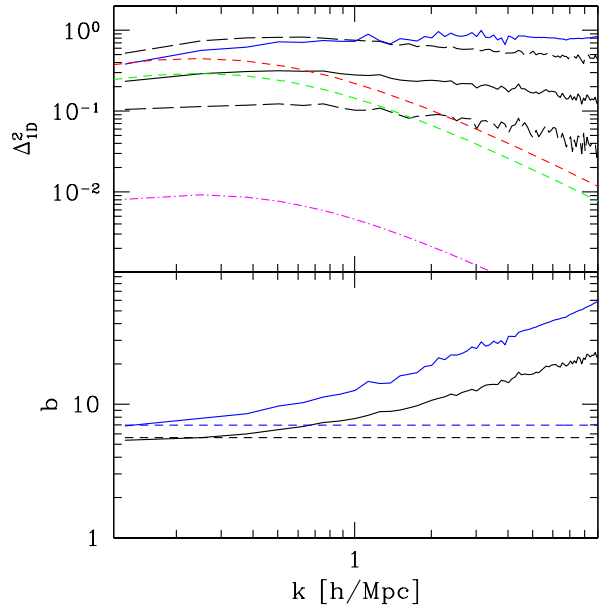}
\caption{\label{fig:bias1D_z10} Same as Figure \ref{fig:bias1D_z6}
except at $z=10$.  The minimum Lyman-$\alpha$ luminosity of
$L_{\rm L\alpha,min}=10^{40}\,(10^{41})\,\rm{ergs/s}$ corresponds to
$M_{\rm halo}=8.2\times10^8\,(3.2\times10^9)\,\msun$ at this redshift.  }
\end{center}
\end{figure}

When analyzing the 1D power spectrum, we do not attempt to produce
mock surveys viewed along the light-cone since we wish to study the
bias of the halos hosting high-redshift galaxies and compare our
results to analytic, linear theory.  Rather, we consider snapshots at $z=6$ and $z=10$.  For this purpose, it is
preferrable that the evolution of the halo mass and correlation
functions will not introduce features in the power spectrum
\citep{ML08b}.  Additionally, stitching together simulation slices to
produce a view along the light-cone would unnecessarily add or
subtract (depending on the prescription used) power from the spectrum
on scales larger than the size of the simulation box and on the scale
of the separation between simulation slices.  As shown in Figures
\ref{fig:bias1D_z6} and \ref{fig:bias1D_z10}, the
$143\,\rm{Mpc}$ comoving size of the simulation box is
sufficiently large to reproduce the linear results on the largest
scales.  We plot our results only for values of $k$ larger than $0.1\,\mpc^{-1}$ to avoid confusion from power loss due to the finite box size.

We calculate the spectrum in terms of real-space wave-numbers, as
opposed to ones in redshift-space, and ignore contamination of
distance measurements by peculiar velocities \citep{Monaco05} 
since they are expected to be small in the early universe.  The redshifts of 
LAEs can be constrained to within a few hundred \rm{km/s}, corresponding 
to a wave-number of $k\sim3\,\rm{Mpc^{-1}}$, while the distances to
high-redshift LBGs can be uncertain to hundreds of comoving \rm{Mpc}
\citep{BI06}.

We compare the power spectrum calculated from the simulation using a
1D Fast Fourier Transform (FFT) to that determined analytically from
linear perturbation theory.  The theoretical 1D power spectrum is
obtained by integrating the 3-dimensional power spectrum over a window
function corresponding to the field-of-view \citep{KP91}:
\begin{equation}\label{eq:power1D}
P_{\rm 1D}(k_w,M)=\frac{(b_{eff}(M)\,D(z))^2}{(2\,\pi)^2}\int\,\,P(k)\,W^2_{xy}\,dk_x\,dk_y,
\end{equation}
where $M$ is the minimum halo mass of the LBGs under consideration,
$k=\sqrt{k_x^2+k_y^2+k_w^2}$, $W_{xy}=W(k_x)\,W(k_y)$, and $W(k_i)$ is
the same as in equation~\ref{eq:window}.  The power measured in the
simulation, however, includes an additional, scale-independent
component from Poisson noise.  This, too, can be calculated
analytically as
\begin{equation}
P_{\rm 1D,Poisson}(M)=(\bar{n}(>M)\,a_x\,a_y\,a_w\,Q(a_w))^{-1}
\end{equation}
and subtracted off, where $\bar{n}(>M)$ is the ST halo mass function of LAE hosts with mass greater than $M$ and
\begin{equation}\label{eq:Q}
Q(a_w)=\frac{1}{2\,\pi}\int_{-\infinity}^{\infinity}\,\,W^2(k_w)\,dk_w.
\end{equation}
Here, $a_z$ can be any length scale and represents the depth of the skewer for which $\sigma^2_P=\left<N\left(>M\right)\right>=\bar{n}\left(>M\right)\,a_x\,a_y\,a_w$ is the Poisson variance in the number of LAEs.  Since,
\begin{equation}
\int_{-\infinity}^{\infinity}\,\,W^2(k_w)\,dk_w\sim2/a_w, \nonumber
\end{equation}
$Q(a_w)\propto1/a_w$, and so $P_{\rm 1D,Poisson}(M)$ is approximately
independent of $a_w$, as expected.  From equation~\ref{eq:power1D}, we can see that the 1D power spectrum at a given scale is a complicated convolution of shorter wavelength modes with the window function.  This integration washes out any features that may have existed at small scales.  Thus, we expect the 1D spectrum to be relatively featureless and smooth.

We analyze the power spectrum for as many skewers through our simulation box that can be packed into the $143\,\mpc$ size, $17^2$ and $15^2$ for $z=6$ and $z=10$, respectively, before rotating the box to a new orientation and collecting additional sets of skewers.  As in \S \ref{sec:pdist}, this scheme is equivalent to choosing an infinite number of skewers each placed randomly on the face of the simulation box.  The argument also applies to additional skewers gained through rotation of the simulation volume.

Figures \ref{fig:bias1D_z6} and \ref{fig:bias1D_z10} compare analytic
and simulation calculations of the 1D power spectrum and bias after
cosmic reionization is assumed to be completed ($z=6$) and in its
early stages at $z=10$, respectively.  In both figures, the black (blue)
lines represent values for a minimum Lyman-$\alpha$ luminosity of
$L_{\rm L\!\alpha,min}=10^{40}\,(10^{41})\,\rm{ergs/s}$ corresponding to
$M_{\rm halo}=1.6\times10^9\,(6.4\times10^9)\,\msun$ at $z=6$ and
$M_{\rm halo}=8.2\times10^8\,(3.2\times10^9)\,\msun$ at $z=10$.  While
LAEs are currently detected to approximately this luminosity at $z=6$,
future surveys should be able to probe down to the same luminosity at
$z=10$.  Solid lines show the average log of the Poisson
subtracted power spectrum (represented as $\Delta^2_{\rm 1D}(k)=k\,P_{\rm 1D}(k)/\pi$)
measured from $3\times17^2$ and $3\times15^2$ skewers for $z=6$ and
$z=10$, respectively.  The factor of $3$ comes from additional
independent skewers we obtained from rotating the simulation box.

Long-dashed lines indicate the estimated $1-\sigma$ variation in the
log of the amplitude of the power spectrum from skewer-to-skewer.  For
clarity, these lines are only shown for
$L_{\rm L\,\alpha,min}=10^{40}\,\rm{ergs/s}$, but they are about the same
size for $L_{\rm L\,\alpha,min}=10^{41}\,\rm{ergs/s}$.  An estimate of the
power from a single skewer will fall within the $1-\sigma$ bounds $68$
percent of the time.  The standard deviation of the log of the power
is approximately the same for all redshift and luminosity threshold
combinations we considered with $\sigma\approx0.4$; it is also
approximately independent of scale.  The precision to which one is able to recover the true mean power spectrum is given by the standard error: $\sigma_e=\sigma/\sqrt{N_{skewers}}$.  Assuming
that our estimate of the standard deviation is close to the true
value and given
our number of independent skewers, we were able to estimate the mean of the log of the power to within
$\sigma_e\approx0.02$.

The level of error in measuring the power informs whether the 1D power
contains sufficient information to distinguish between galaxies of
different masses and luminosities.  This differentiability is
important in constraining the mass-luminosity relationship and happens
when the standard errors for each power spectrum using $N_{\rm skewers}$ independent fields-of-view (in this case all having dimensions of
$3.4'\times3.4'$) are small enough that they don't overlap:
$\left|\mu_{41}-\mu_{40}\right|=(\sigma_{41}+\sigma_{40})/\sqrt{N_{\rm skewers}}$,
where $\mu_x$ and $\sigma_x$ are the mean and standard deviations of
the log of the power for the luminosity thresholds corresponding to
$L_{\rm L\,\alpha,min}=10^x \rm{ergs/s}$.  
On the largest scales probed, the mean of the log of the power differs by 
$\approx0.23$ 
for both redshifts (with a weak 
scale 
dependence) 
and $\sigma_{41}\approx\sigma_{40}\approx0.4$.  
Thus, approximately 12 independent skewers are needed to distinguish between the two luminosity thresholds based on their 1D power spectrum.  

The short-dashed green (red) line shows the analytic power-spectrum
according to linear perturbation theory, with the same luminosity
threshold as the black (blue) line.  With no fit other than the
calculations described above, the analytic predictions match the data
from the simulations quite well on the largest scales.  As soon as
nonlinearities dominate on small scales, the simulation results and
the analytic calculations diverge in their predictions.  However, it is important to note that the nonlinearities at work are on even smaller scales than the deviation scale due to aliasing.

\begin{figure}
\begin{center}
\includegraphics[width=\columnwidth]{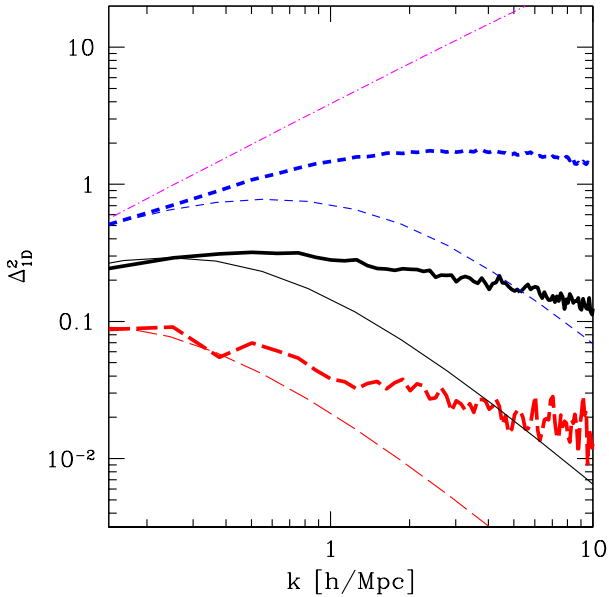}
\caption{\label{fig:bias1D_z10_sizes} A comparison of the 1D power spectrum of halos hosting LAEs, as in Fig. \ref{fig:bias1D_z10}, in surveys with different fields-of-view at $z=10$.  We set $L_{\rm L\alpha,min}=10^{40}\,\rm{ergs/s}$ for each survey.  Thick lines show the simulated average log of the Poisson-subtracted, dimensionless power spectrum, while thin ones represent calculations from linear theory.  Solid, black lines correspond to a $3.4'\times3.4'$ field-of-view, while the short-dashed, blue and long-dashed, red lines denote $1'\times1'$ and $10'\times10'$ fields-of-view, respectively.  The field-to-field variance from Figures \ref{fig:bias1D_z6} and \ref{fig:bias1D_z10} have been omitted for clarity.  For reference, we have plotted the linear theory spectrum for the dark matter assuming a pure 1D survey with an infinitesimal field-of-view as the dotted-dashed, magenta curve.
}
\end{center}
\end{figure}

In the bottom panels of Figures \ref{fig:bias1D_z6} and
\ref{fig:bias1D_z10}, we show the halo bias,
\begin{equation}\label{eq:bias}
b=\sqrt{P^{\rm halo}_{\rm 1D,halo}/P^{\rm analytic}_{\rm 1D,dm}},
\end{equation}
measured from the simulation (solid lines) and compare the results to
the analytic ST calculation (short-dashed lines).  Here,
$P^{\rm analytic}_{\rm 1D,dm}$ is the linear, analytic calculation of the 1D
dark matter power from equation~(\ref{eq:power1D}) with
$b_{\rm eff}(M)=1$.  These plots show the same trend as in the upper
panels with increasing divergence between simulation and analytic
calculations on small scales.  This is precisely the result obtained
from the full three-dimensional power by \citet{TC07}.  However, note
that the linear and nonlinear values do not match up exactly for
$L_{\rm L\,\alpha,min}=10^{41}\,\rm{ergs/s}$ at $z=6$.  The ST mass
function is not a perfect fit over all masses and so it results in a
slight deviation in the derived bias.  This effect also manifests
itself in the upper panel as a small deviation on large scales between
the simulation and analytic power for the same luminosity and
redshift.

A striking aspect of the 1D power spectrum illustrated in the top panels is that, for the $3.4'\times3.4'$ field-of-view, it remains relatively flat and featureless over a range of $k$-values spanning an order-of-magnitude or more.  This smoothness is expected since both the aliasing of small-scale power and the averaging of many skewers wash out small features in the spectrum.  However, the flatness is unique to our choice of field-of-view.  

In Figure \ref{fig:bias1D_z10_sizes}, we investigate how observational results from different fields-of-view can been compared by plotting the 1D power simulated in $3.4'\times3.4'$ pencil-beam skewers alongside those using $1'\times1'$ and $10'\times10'$ fields-of-view.  The behavior in the interesting range of modes differs among the three cases because of the different modes allowed in the integration by the window function, which mixes up modes in a way that cannot be easily de-convolved (see Eq. \ref{eq:power1D}).  

Comparing the the linear theory predictions, we see that the position of the turn-over is directly related to the size of the window function with the narrower field-of-view turning over at larger $k$.  Although the simulated position of the turn-over still increases monotonically with field-of-view size, the aliasing of nonlinear small-scale power changes the window function convolution in such a way that the position of the peak is not easily determined from the linear case.  

It is also clear that the flatness we noted in the simulated spectrum for the $3.4'\times3.4'$ field-of-view is not generic; the spectrum for the narrowest field-of-view first rises with $k$ and then decreases.  The $10'\times10'$ field-of-view produces a spectrum that is flatter than the one produced by the $1'\times1'$ field-of-view, but since the peak is also at much larger scales, we end up probing the spectrum in the regime where it begins to fall off at small scales.  The extra bumps and wiggles in the curve result from fewer skewers of that size being available for averaging.  

For reference, we have also shown in Figure \ref{fig:bias1D_z10_sizes} the 1D dark matter power spectrum in an idealized 1D survey with an infinitesimal field-of-view.  The aliased power in this case increases the amplitude of the spectrum even more than the bias factor in the LAE spectra using realistic fields-of-view.  The figure shows the 1D power approaching this limit as the field-of-view for realistic surveys decreases.

\section{Conclusions}\label{sec:conclusions}

In this work we have analyzed the statistics of galaxies in mock
``pencil beam" surveys of high-redshift LBGs and LAEs with narrow
fields-of-view produced using N-body simulations.  We computed the
effect of sample variance and nonlinear bias and showed how these will
influence measurements of galaxy counts in existing and upcoming
surveys.

While the variance due to fluctuations in the local overdensity in LBG
surveys increases almost monotonically with increasing luminosity (or
halo mass), it dominates only at low luminosities and is overtaken at
high luminosities by a Poisson variance that rises more steeply as the
average number of galaxies in the survey drops.  However, the
overdensities become nonlinear on small scales.  A given point is more
likely to be in underdense regions than in overdense regions because
of their larger volume filling fraction. The resulting skewness of the
probability distribution of galaxy counts creates slightly more
uncertainty in obtaining the true number density of galaxies at the
observed redshift than would otherwise be expected.  This can be seen
graphically in Fig. \ref{fig:pdist8-10} where the simulation histogram
is most closely matched by a skewed, log-normal distribution whose
peak is centered on negative values of the overdensity.  The skewness
of the distribution and the resulting increased probability of
surveying an underdense region should be taken into account when
interpreting galaxy counts in the context of the ionizing UV
background they produce at high redshift, or else it would result in
an underestimate of the background.  Precise measurements of the cosmic variance in a survey is also important for understanding the differences between two independent samples.  Our results and methods can be used to calibrate these errors for different fields-of-view and survey depths appropriate to a given study instead of relying only on linear theory estimates at a single redshift as in \citet{Bouwens08a}.

Whether or not future surveys of LAEs will have the angular resolution
to compute angular correlation functions in the plane of the sky, they
will probe many more modes along the line-of-sight.  We expect the 1D
power spectrum from surveys with a JWST-sized field-of-view to exhibit a roughly flat, smooth profile.  Aliasing and the averaging of many skewers are expected to result in a relatively featureless spectrum unable to probe void structure like that seen in \citet{Broadhurst90}.  The flatness, on the other hand, is unique to the particular choice of field-of-view, and comparing JWST results, such as measurements of the galaxy bias, with those from other surveys will not be trivial due to the complicated convolution with the window function.  Nonetheless, if the survey parameters are known, the amplitude on the largest scales matches well with the
expectation from linear theory.  However, on small scales, the
nonlinear bias becomes large.  Given their limited fields-of-view,
a survey with JWST will need tens of pointings to be able to distinguish
between host halo masses different by an order-of-magnitude limiting
its usefulness in determining their mass-to-light ratios.  Finally,
our calculations ignored the complex effects of reionization on the
amplitude of the power spectrum.  The mass-to-light ratio for LAEs at
redshifts during reionization could first be calibrated by
extrapolating determinations made after reionization, through either
clustering or abundance matching.  With this information in hand,
differences between our results and observations will show the effect
of Lyman-$\alpha$ transmission through the HII regions around ionizing
sources.  However, probing reionization through the changing amplitude of the 1D spectrum \citep{McQuinn07} will require particular consideration of the field-of-view.

\section{Acknowledgements}

We would like to thank Mark Dijkstra for useful discussions.   HT is supported by an Institute for Theory and Computation Fellowship.  This
research was also supported in part by NASA grants NNX08AL43G and LA and by
Harvard University funds.


\begin{thebibliography}{40}
\expandafter\ifx\csname natexlab\endcsname\relax\def\natexlab#1{#1}\fi

\bibitem[{{Babich} \& {Loeb}(2006)}]{dBL06}
{Babich}, D., \& {Loeb}, A. 2006, \apj, 640, 1

\bibitem[{{Barkana} \& {Loeb}(2001)}]{BL01}
{Barkana}, R., \& {Loeb}, A. 2001, \physrep, 349, 125

\bibitem[{{Bouwens} \& {Illingworth}(2006)}]{BI06}
{Bouwens}, R., \& {Illingworth}, G. 2006, New Astronomy Review, 50, 152

\bibitem[{{Bouwens} {et~al.}(2006){Bouwens}, {Illingworth}, {Blakeslee}, \&
  {Franx}}]{Bouwens06}
{Bouwens}, R., {Illingworth}, G., {Blakeslee}, J., \& {Franx}, M. 2006, \apj,
  653, 53

\bibitem[{{Bouwens} {et~al.}(2009){Bouwens}, {Illingworth}, {Bradley}, {Ford},
  {Franx}, {Zheng}, {Broadhurst}, {Coe}, \& {Jee}}]{Bouwens08b}
{Bouwens}, R.~J., {Illingworth}, G.~D., {Bradley}, L.~D., {Ford}, H., {Franx},
  M., {Zheng}, W., {Broadhurst}, T., {Coe}, D., \& {Jee}, M.~J. 2009, \apj,
  690, 1764

\bibitem[{{Bouwens} {et~al.}(2008)}]{Bouwens08a}
{Bouwens}, R.~J., {et~al.} 2008, \apj, 686, 230

\bibitem[{{Broadhurst} {et~al.}(1990){Broadhurst}, {Ellis}, {Koo}, \&
  {Szalay}}]{Broadhurst90}
{Broadhurst}, T.~J., {Ellis}, R.~S., {Koo}, D.~C., \& {Szalay}, A.~S. 1990,
  \nat, 343, 726

\bibitem[{{Dunkley} {et~al.}(2009){Dunkley}, {Komatsu}, {Nolta}, {Spergel},
  {Larson}, {Hinshaw}, {Page}, {Bennett}, {Gold}, {Jarosik}, {Weiland},
  {Halpern}, {Hill}, {Kogut}, {Limon}, {Meyer}, {Tucker}, {Wollack}, \&
  {Wright}}]{Dunkley09}
{Dunkley}, J., {Komatsu}, E., {Nolta}, M.~R., {Spergel}, D.~N., {Larson}, D.,
  {Hinshaw}, G., {Page}, L., {Bennett}, C.~L., {Gold}, B., {Jarosik}, N.,
  {Weiland}, J.~L., {Halpern}, M., {Hill}, R.~S., {Kogut}, A., {Limon}, M.,
  {Meyer}, S.~S., {Tucker}, G.~S., {Wollack}, E., \& {Wright}, E.~L. 2009,
  \apjs, 180, 306

\bibitem[{{Gnedin} \& {Kravtsov}(2006)}]{GK06}
{Gnedin}, N.~Y., \& {Kravtsov}, A.~V. 2006, \apj, 645, 1054

\bibitem[{{Kaiser} \& {Peacock}(1991)}]{KP91}
{Kaiser}, N., \& {Peacock}, J.~A. 1991, \apj, 379, 482

\bibitem[{{Kobayashi} {et~al.}(2007){Kobayashi}, {Totani}, \&
  {Nagashima}}]{KTN07}
{Kobayashi}, M.~A.~R., {Totani}, T., \& {Nagashima}, M. 2007, \apj, 670, 919

\bibitem[{{Kobayashi} {et~al.}(2009){Kobayashi}, {Totani}, \&
  {Nagashima}}]{KTN09}
---. 2009, arXiv:astro-ph/0902.2882

\bibitem[{{Komatsu} {et~al.}(2009){Komatsu}, {Dunkley}, {Nolta}, {Bennett},
  {Gold}, {Hinshaw}, {Jarosik}, {Larson}, {Limon}, {Page}, {Spergel},
  {Halpern}, {Hill}, {Kogut}, {Meyer}, {Tucker}, {Weiland}, {Wollack}, \&
  {Wright}}]{Komatsu09}
{Komatsu}, E., {Dunkley}, J., {Nolta}, M.~R., {Bennett}, C.~L., {Gold}, B.,
  {Hinshaw}, G., {Jarosik}, N., {Larson}, D., {Limon}, M., {Page}, L.,
  {Spergel}, D.~N., {Halpern}, M., {Hill}, R.~S., {Kogut}, A., {Meyer}, S.~S.,
  {Tucker}, G.~S., {Weiland}, J.~L., {Wollack}, E., \& {Wright}, E.~L. 2009,
  \apjs, 180, 330

\bibitem[{{Kravtsov} \& {Gnedin}(2005)}]{KG05}
{Kravtsov}, A.~V., \& {Gnedin}, O.~Y. 2005, \apj, 623, 650

\bibitem[{{Le Delliou} {et~al.}(2005){Le Delliou}, {Lacey}, {Baugh},
  {Guiderdoni}, {Bacon}, {Courtois}, {Sousbie}, \& {Morris}}]{LeDelliou05}
{Le Delliou}, M., {Lacey}, C., {Baugh}, C.~M., {Guiderdoni}, B., {Bacon}, R.,
  {Courtois}, H., {Sousbie}, T., \& {Morris}, S.~L. 2005, \mnras, 357, L11

\bibitem[{{Le Delliou} {et~al.}(2006){Le Delliou}, {Lacey}, {Baugh}, \&
  {Morris}}]{LeDelliou06}
{Le Delliou}, M., {Lacey}, C.~G., {Baugh}, C.~M., \& {Morris}, S.~L. 2006,
  \mnras, 365, 712

\bibitem[{{Madau} {et~al.}(2008){Madau}, {Kuhlen}, {Diemand}, {Moore}, {Zemp},
  {Potter}, \& {Stadel}}]{Madau08b}
{Madau}, P., {Kuhlen}, M., {Diemand}, J., {Moore}, B., {Zemp}, M., {Potter},
  D., \& {Stadel}, J. 2008, \apjl, 689, L41

\bibitem[{{Madau} {et~al.}(1998){Madau}, {Pozzetti}, \& {Dickinson}}]{MPD98}
{Madau}, P., {Pozzetti}, L., \& {Dickinson}, M. 1998, \apj, 498, 106

\bibitem[{{Matarrese} {et~al.}(1997){Matarrese}, {Coles}, {Lucchin}, \&
  {Moscardini}}]{Matarrese97}
{Matarrese}, S., {Coles}, P., {Lucchin}, F., \& {Moscardini}, L. 1997, \mnras,
  286, 115

\bibitem[{{McQuinn} {et~al.}(2007){McQuinn}, {Hernquist}, {Zaldarriaga}, \&
  {Dutta}}]{McQuinn07}
{McQuinn}, M., {Hernquist}, L., {Zaldarriaga}, M., \& {Dutta}, S. 2007, \mnras,
  381, 75

\bibitem[{{Mobasher} {et~al.}(2005){Mobasher}, {Dickinson}, {Ferguson},
  {Giavalisco}, {Wiklind}, {Stark}, {Ellis}, {Fall}, {Grogin}, {Moustakas},
  {Panagia}, {Sosey}, {Stiavelli}, {Bergeron}, {Casertano}, {Ingraham},
  {Koekemoer}, {Labb{\'e}}, {Livio}, {Rodgers}, {Scarlata}, {Vernet},
  {Renzini}, {Rosati}, {Kuntschner}, {K{\"u}mmel}, {Walsh}, {Chary},
  {Eisenhardt}, {Pirzkal}, \& {Stern}}]{Mobasher05}
{Mobasher}, B., {Dickinson}, M., {Ferguson}, H.~C., {Giavalisco}, M.,
  {Wiklind}, T., {Stark}, D., {Ellis}, R.~S., {Fall}, S.~M., {Grogin}, N.~A.,
  {Moustakas}, L.~A., {Panagia}, N., {Sosey}, M., {Stiavelli}, M., {Bergeron},
  E., {Casertano}, S., {Ingraham}, P., {Koekemoer}, A., {Labb{\'e}}, I.,
  {Livio}, M., {Rodgers}, B., {Scarlata}, C., {Vernet}, J., {Renzini}, A.,
  {Rosati}, P., {Kuntschner}, H., {K{\"u}mmel}, M., {Walsh}, J.~R., {Chary},
  R., {Eisenhardt}, P., {Pirzkal}, N., \& {Stern}, D. 2005, \apj, 635, 832

\bibitem[{{Monaco} {et~al.}(2005){Monaco}, {M{\o}ller}, {Fynbo}, {Weidinger},
  {Ledoux}, \& {Theuns}}]{Monaco05}
{Monaco}, P., {M{\o}ller}, P., {Fynbo}, J.~P.~U., {Weidinger}, M., {Ledoux},
  C., \& {Theuns}, T. 2005, \aap, 440, 799

\bibitem[{{Moore} {et~al.}(2006){Moore}, {Diemand}, {Madau}, {Zemp}, \&
  {Stadel}}]{Moore06}
{Moore}, B., {Diemand}, J., {Madau}, P., {Zemp}, M., \& {Stadel}, J. 2006,
  \mnras, 368, 563

\bibitem[{{Mu{\~n}oz} \& {Loeb}(2008)}]{ML08b}
{Mu{\~n}oz}, J.~A., \& {Loeb}, A. 2008, \mnras, 386, 2323

\bibitem[{{Mu{\~n}oz} {et~al.}(2009){Mu{\~n}oz}, {Madau}, {Loeb}, \&
  {Diemand}}]{Munoz09}
{Mu{\~n}oz}, J.~A., {Madau}, P., {Loeb}, A., \& {Diemand}, J. 2009, \mnras,
  400, 1593

\bibitem[{{Nagamine} {et~al.}(2008){Nagamine}, {Ouchi}, {Springel}, \&
  {Hernquist}}]{Nagamine08}
{Nagamine}, K., {Ouchi}, M., {Springel}, V., \& {Hernquist}, L. 2008,
  arXiv:astro-ph/0802.0228

\bibitem[{{Orsi} {et~al.}(2008){Orsi}, {Lacey}, {Baugh}, \& {Infante}}]{Orsi08}
{Orsi}, A., {Lacey}, C.~G., {Baugh}, C.~M., \& {Infante}, L. 2008, \mnras, 391,
  1589

\bibitem[{{Overzier} {et~al.}(2009){Overzier}, {Guo}, {Kauffmann}, {De Lucia},
  {Bouwens}, \& {Lemson}}]{Overzier09}
{Overzier}, R.~A., {Guo}, Q., {Kauffmann}, G., {De Lucia}, G., {Bouwens}, R.,
  \& {Lemson}, G. 2009, \mnras, 394, 577

\bibitem[{{Richard} {et~al.}(2008){Richard}, {Stark}, {Ellis}, {George},
  {Egami}, {Kneib}, \& {Smith}}]{Richard08}
{Richard}, J., {Stark}, D.~P., {Ellis}, R.~S., {George}, M.~R., {Egami}, E.,
  {Kneib}, J.-P., \& {Smith}, G.~P. 2008, \apj, 685, 705

\bibitem[{{Sheth} {et~al.}(2001){Sheth}, {Mo}, \& {Tormen}}]{SMT01}
{Sheth}, R.~K., {Mo}, H.~J., \& {Tormen}, G. 2001, \mnras, 323, 1

\bibitem[{{Sheth} \& {Tormen}(1999)}]{ST99}
{Sheth}, R.~K., \& {Tormen}, G. 1999, \mnras, 308, 119

\bibitem[{{Somerville} {et~al.}(2004){Somerville}, {Lee}, {Ferguson},
  {Gardner}, {Moustakas}, \& {Giavalisco}}]{Somerville04}
{Somerville}, R.~S., {Lee}, K., {Ferguson}, H.~C., {Gardner}, J.~P.,
  {Moustakas}, L.~A., \& {Giavalisco}, M. 2004, \apjl, 600, L171

\bibitem[{{Stark} {et~al.}(2007{\natexlab{a}}){Stark}, {Ellis}, {Richard},
  {Kneib}, {Smith}, \& {Santos}}]{Stark07}
{Stark}, D.~P., {Ellis}, R.~S., {Richard}, J., {Kneib}, J.-P., {Smith}, G.~P.,
  \& {Santos}, M.~R. 2007{\natexlab{a}}, \apj, 663, 10

\bibitem[{{Stark} {et~al.}(2007{\natexlab{b}}){Stark}, {Loeb}, \&
  {Ellis}}]{SLE07}
{Stark}, D.~P., {Loeb}, A., \& {Ellis}, R.~S. 2007{\natexlab{b}}, \apj, 668,
  627

\bibitem[{{Trac} \& {Cen}(2007)}]{TC07}
{Trac}, H., \& {Cen}, R. 2007, \apj, 671, 1

\bibitem[{{Trac} {et~al.}(2008){Trac}, {Cen}, \& {Loeb}}]{Trac08}
{Trac}, H., {Cen}, R., \& {Loeb}, A. 2008, \apjl, 689, L81

\bibitem[{{Trenti} \& {Stiavelli}(2008)}]{TS08}
{Trenti}, M., \& {Stiavelli}, M. 2008, \apj, 676, 767

\bibitem[{{Wiklind} {et~al.}(2008)}]{Wiklind08}
{Wiklind}, T., {et~al.} 2008, \apj, 676, 781

\bibitem[{{Wyithe} \& {Loeb}(2006)}]{WL06}
{Wyithe}, J.~S.~B., \& {Loeb}, A. 2006, \nat, 441, 322

\bibitem[{{Wyithe} \& {Loeb}(2007)}]{WL07}
---. 2007, \mnras, 382, 921

\end{thebibliography}
\end{document}